\documentclass[%
 reprint,
 nofootinbib,
 amsmath,amssymb,
 aps,
]{revtex4-1}
\pdfoutput=1
\usepackage{graphicx}
\usepackage{dcolumn}
\usepackage{tikz}
\usepackage{bm}
\usepackage{pgfplots}
\usepackage{pgfplotstable}
\pgfplotsset{compat=newest}
\usepackage{siunitx}
\usepackage{graphicx}
\usepackage{ulem}
\usepackage{enumerate}

\newcommand{\beq}{\begin{equation}}
\newcommand{\eeq}{\end{equation}}
\newcommand{\bea}{\begin{eqnarray}}
\newcommand{\eea}{\end{eqnarray}}
\allowdisplaybreaks

\begin{document}


\title{KK Gravitons at LHC2}

\author{Barry M. Dillon}
\email{barry.dillon@plymouth.ac.uk}
\email{b.dillon@sussex.ac.uk}
\affiliation{Centre for Mathematical Sciences, Plymouth University, PL4 8AA, UK}
\affiliation{Department of Physics and Astronomy, University of Sussex, Brighton, United Kingdom}
\author{Veronica Sanz}
\email{v.sanz@sussex.ac.uk}
\affiliation{Department of Physics and Astronomy, University of Sussex, Brighton, United Kingdom}




\date{\today}

\begin{abstract}
In this work we study constraints from new searches for heavy particles at the LHC on the allowed masses and couplings of a KK Graviton in a holographic composite Higgs model.  
Keeping new electroweak states heavy such that electroweak precision tests are satisfied, we control the mass of the lightest KK graviton using a brane kinetic term.
With this we study KK graviton masses from $0.5-3$ TeV.
In our analysis we also employ {\it Little Randall-Sundrum} (RS) Models, characterised by a lower UV scale in the 5D model which in turn implies modified couplings to massless bulk fields.
Viewing this scenario as a strongly coupled 4D theory with a composite Higgs boson, the KK graviton is interpreted as a composite spin-2 state and the varying UV scale corresponds to a varying intermediate scale between the cutoff of the low energy effective theory and the Planck scale.
We find that KK gravitons with masses in the range $[500,3000]$ GeV are compatible with current collider constraints, where the most promising channels for detecting these states are the di-photon and $ZZ$ channels.
A detection is more likely in the little RS models, in which the dual gauge theory has a larger number of colours than in traditional RS models.
\end{abstract}

\pacs{Valid PACS appear here}
\maketitle


\section{Introduction}

Run 2 of the LHC has yet to discover physics beyond the Standard Model (SM).
Early hints of new physics were uncovered in the form of a scalar resonance in the di-photon channel, however these have since been shown to have arisen from a statistical fluctuation in the data.
Much work was done based on these early results \cite{Franceschini:2015kwy,Ellis:2015oso,DiChiara:2015vdm,Chakrabortty:2015hff,Gupta:2015zzs,Staub:2016dxq,Sanz:2016auj,Kim:2015ksf,Martini:2016ahj,Geng:2016xin,Ahmed:2015uqt,Cox:2015ckc,Davoudiasl:2015cuo,Boos:2016ytd,Han:2015cty,Falkowski:2016glr,Arun:2015ubr,Liu:2016mpd,Bellazzini:2015nxw,Harigaya:2015ezk,Davoudiasl:2015cuo,No:2015bsn,Carmona:2016jhr,Hewett:2016omf,Han:2015cty}, and the work we present here was also inspired by attempts at explaining this di-photon resonance.

The absence of a new physics discovery at the LHC prompts questions about the scale, if any, of physics which gives a dynamical explanation of electroweak symmetry breaking (EWSB) and an explanation of the flavour hierarchies in the SM.
Composite Higgs models present an attractive framework in which we can answer these questions \cite{Terazawa:1976xx,Terazawa:1979pj,Kaplan:1983sm,Georgi:1984af,Dugan:1984hq} , however the models predict new physics at scales not far above the electroweak (EW) scale, i.e. $\sim1$ TeV.
The smoking gun signal of a composite Higgs model is widely thought to be a light spin-1/2 top partner state, however spin-1 and spin-2 states, as well as deviations in the Higgs couplings, are also expected.
New physics in flavour observables or electroweak precision observables (EWPOs) may also be expected, however these effects can be sufficiently reduced by imposing additional global symmetries on the new physics.

To obtain quantitative predictions for this strongly coupled scenario we make use of the five-dimensional (5D) holographic models, these are proposed to be dual to classes of strongly coupled gauge theories in 4D. In this framework, geometries in 5D are mapped to 4D theories with different running behaviour, exhibiting various types of symmetries.  Among those choices of geometries, the Randall-Sundrum (RS) is particularly interesting, as it maps to 4D theories with some degree of conformal invariance. In this set-up, UV boundary conditions of the 4D running and spontaneous breaking of conformal invariance are mapped to  two 3-branes at the boundaries of the 5D bulk in which a negative cosmological constant generates an AdS background \cite{Randall:1999ee,Goldberger:1999uk,Gherghetta:2000qt,Davoudiasl:1999tf,delAguila:2003bh,Huber:2000ie}.  Fluctuations of this 5D background can be described by two types of particles, one is the radion (spin-0), and the other is a tower of KK gravitons (spin-2).
The 5D models predict that the physics of the radion and KK gravitons are largely independent of the global symmetry structure of any specific composite Higgs model, this is not the case for the spin-1 and spin-1/2 states.

Irrespective of the spin, in RS the resonances are linked to the origin of electroweak symmetry breaking, and hence the Higgs sector.
They are precisely identified with composite resonances of the dual 4D strongly coupled gauge theory.
There has been a fair amount of work exploring this relation, and the scenario which stands out as most natural is the {\it holographic composite Higgs}. This model is realised with the Standard Model fields propagating in the 5D bulk, and the Higgs as the fifth components of some 5D gauge field. The extra-dimensional (or holographic) Higgs is then dual to a pseudo-Goldstone boson arising from the breaking of an approximate global symmetry due to the strong dynamics~\cite{Contino:2003ve,Agashe:2004rs,Agashe:2006at,Contino:2006qr,Hirn:2006wg,Sanz:2015sua,Croon:2015wba}.  

In this paper we will study experimental bounds on the parameter space of a KK graviton arising from a 5D model, where the 5D properties of the Higgs are fixed by the requirement that its dual interpretation is that of a composite pseudo-Goldstone boson state.
One thing that makes this study different from most other phenomenological studies of KK gravitons is that we study a large range of `5D volumes'.
In the dual 4D gauge theory the 5D volume is inversely related to the number of colours in the confining gauge theory, more details of this are given in the main text. 
AdS models with small 5D volumes have been studied previously and are known as `Little Randall-Sundrum models' \cite{Davoudiasl:2008hx,McDonald:2008ss,Davoudiasl:2008nr,Bauer:2008xb,Davoudiasl:2009jk,Davoudiasl:2010fb,George:2011sw,Croon:2015wba}.  In particular in \cite{Croon:2015wba} it has been shown that a lower UV scale in the Composite Higgs scenarios results in less fine-tuning in the Higgs potential and allows for top partners to be naturally be heavier than $1$ TeV.  Note that while performing this analysis we assume that the radion is heavy enough such that it does not significantly affect the phenomenology of the KK graviton. 

In the first section we will review some of the work on KK gravitons, in particular; how the first graviton KK mode can be made lighter than other KK states in the model, the role of the 5D volume, and how the couplings of the KK graviton to the SM are determined.  We then move on to discuss the relationship between these 5D models and the dual 4D gauge theories.  In the next section we present expressions for the branching fractions of the KK graviton to the SM states, and plot the production cross-section of the KK graviton as a function of its mass.  In the last two sections we present the experimental bounds on the model and discuss the phenomenology of the KK graviton.

\section{Warped KK Gravitons}

Let us start by describing the general properties of warped gravitons in RS. In deriving the couplings and 5D properties of the KK gravitons we closely follow the work in \cite{Davoudiasl:1999jd,Agashe:2007zd,Fitzpatrick:2007qr,Lee:2014caa,Lee:2013bua,Falkowski:2016glr}.  To begin, let us consider the following 5D action,
\begin{eqnarray}
S_5=\int d^4x\int_0^L dy~\sqrt{|g|} M_{*}^3\left(-\frac{\mathcal{R}_5}{2}+6k^2 \right) \nonumber\\
+\left( \sqrt{|g|}M_{*}^3(k_B\pm \frac{1}{2 k}r_{0,L}\mathcal{R}_4)\right)\Big\vert^L_0.
\label{Sact}
\end{eqnarray}
where the RS metric and its fluctuations can be described by,
\begin{equation}
\label{RSds}
ds^2=e^{-2ky}\left(\eta_{\mu\nu}+h_{\mu\nu}(x,y)\right)dx^{\mu}dx^{\nu} - dy^2.
\end{equation}
The co-ordinate $y$ labels the position along the extra dimension, which is bounded by two 3-branes at $y=0$ (UV) and $y=L$ (IR). 

The quantity $k$ is known as the curvature constant and parametrises the warping in the bulk of the extra dimension, and $M_{*}$ is the UV mass scale in the 5D theory.  The $h_{\mu\nu}(x,y)$ fluctuation would correspond to the bulk graviton field.  Note that we have neglected fluctuations along the $y$ direction, which would represent the radion dynamics.

One can perform a KK decomposition on the field $h_{\mu\nu} (x,y) = \sum_n f^g_n (y) h_{\mu\nu}^n(x) $, where each Kaluza-Klein mode $h_{\mu\nu}^n(x)$ represents a 4D massive graviton of mass $m_n$ with a 5D profile $f^g_n$ obeying the following eigenvalue equation in the bulk,
\begin{eqnarray}
\label{bulkGeom}
\partial_y^2f^g_n-4k\partial_yf^g_n+m_n^2e^{2ky}f^g_n=0 ,
\end{eqnarray}
where the mass $m_n$ of the $n^{th}$ KK mode is of the order $M_{KK} \equiv ke^{-kL}$.  

Turning to the boundary terms in Eq.~\ref{Sact}, $k_B$ is a brane tension, whose effect is to compensate the negative bulk cosmological constant $k$. The Ricci scalar terms on the branes proportional to $r_{0,L}$ imply brane kinetic terms (BKTs) for the graviton modes (and the radion's). These BKTs result in modifications to the boundary conditions of the 5D profiles of on-shell 4D graviton modes,
\begin{align}
\label{bulkGbcs}
    \left(k \partial_yf^g_n+r_0m_n^2f^g_n\right)\Big\vert_0&=0   \\
     \left(e^{-2kL} k \partial_yf^g_n-r_Lm_n^2f^g_n\right)\Big\vert_L&=0. 
\end{align}

These boundary conditions along with the eigenvalue equation permit a flat massless graviton zero mode. 
The general solution to eq.~\ref{bulkGeom} is,
\begin{equation}
\label{bulkGsol1 }
f^g_n=\frac{e^{2ky}}{N_n^{1/2}}\left(J_2\left(z_ne^{k(y-L)}\right)+\alpha_nY_2\left(z_ne^{k(y-L)}\right)\right)
\end{equation}
where $z_n=\frac{m_n}{M_{KK}}$.  The Bessel functions are periodic in their arguments thus we see that these wave functions are exponentially localised towards the IR. 
The value of $\alpha_n$ and $m_n$ is fixed by the boundary conditions and $N_n$ by the normalisation condition.   Applying the UV boundary conditions we find,
\bea
\alpha_n^{(i)} =-\frac{J_1\left( z'_n \right)+r_0 z'_nJ_2\left( z'_n\right)}{Y_1\left(z'_n\right)+r_0 z'_nY_2\left(z'_n\right)}
\eea
where $z'_n= m_n/k=e^{-k L} z_n$.
Due to the IR localisation of the KK gravitons the UV boundary condition will not be important for the physics we study.
Also note that when $m_n\neq 0$ a large UV BKT has a similar effect as a localised UV mass term for the KK modes, whereas it has no effect on the profile of the massless zero mode.  

The values of the masses,  $m_n$, are fixed by applying the IR boundary condition,
\bea
J_1\left(z_n\right)-r_L z_n J_2\left(z_n\right) = - \alpha_n^{(i)}  \left( Y_1\left(z_n\right)-r_L  z_n Y_2\left(z_n\right) \right) \ ,
\eea
where we have dropped additional terms suppressed by $m_n/k$.  
The $Y_a(x)$ function diverges at $x\rightarrow 0$ hence the terms $\sim Y_1\left(z'_n\right)$ dominate.  With no IR BKT, the masses of the lowest lying modes are then approximately given by the zeroes of $J_1\left(z_n\right)$, namely 3.8, 7, $10.2$, $13.3$ in units of $M_{KK}$.  In contrast, the KK masses of bulk spin-1 fields are given by the zeroes of $J_0\left(z_n\right)$, i.e.  2.4, 5.5, 8.7 and so on.  This pattern can be altered with non-zero BKTs for either the graviton or the spin-1 fields.
In our work we are specifically interested in obtaining light spin-2 states, thus we assume no BKTs for the spin-1 fields.
We see that the lightest KK graviton mass drops steeply with small values of $r_L$, reaching $\sim 1.7~M_{KK}$ for $r_L=1$.

The $r_L=1$ limit is important, since above this value the radion fluctuation becomes ghost-like, signalling an instability in the model\footnote{We would like to thank Kaustubh Agashe for bringing this to our attention, and discussing the issue in detail.}\cite{Davoudiasl:2003zt,Falkowski:2016glr}.
There has been some discussion on this issue and viable scenarios in which $r_L>1$ have been described \cite{Dillon:2016bsb}.
It is also worth mentioning that in deconstructed models of warped extra dimensions we would find the same graviton spectrum and features while the radion state would not be present and thus the $r_L$ bound would be irrelevant \cite{ArkaniHamed:2001ca}. 
Therefore to be as general as possible we study regions of parameter space both above and below the $r_L=1$ bound. 

\begin{figure}[h!]
  \begin{center}
\includegraphics[scale=0.38]{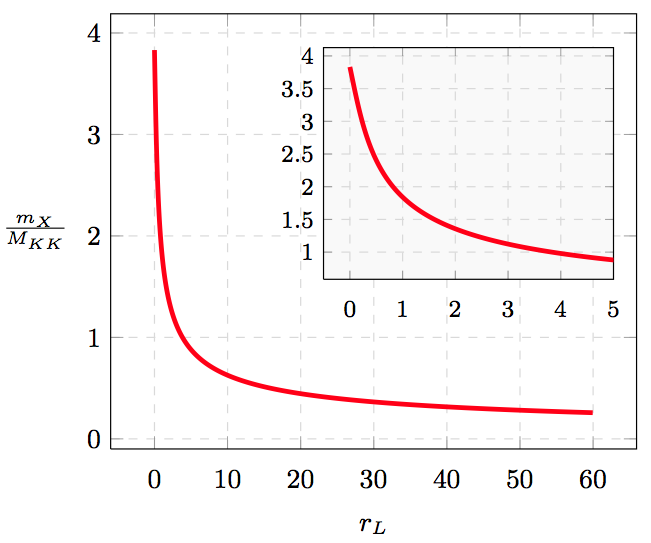}
    \caption{The mass of the lightest KK graviton as a function of the infra-red BKT. \label{Xmass}}
  \end{center}
\end{figure}

For large values of $r_L$, one can expand the Bessel functions in the IR boundary condition and obtain the following approximate solution for the lightest massive mode,
\begin{equation}
\label{mxrl}
m_X\simeq\frac{2M_{KK}}{\sqrt{r_L}}.
\end{equation}      
Thus for increasing values of $r_L$, one can suppress the lightest spin-2 mode and make it lighter than the spin-1 modes.  
A more accurate relationship between $m_X$ and $r_L$ can only be determined numerically, we have computed this and plotted the relationship in figure \ref{Xmass}.

\subsection{The UV scale \& little RS}

To completely fix the integration constants in the 5D wave functions we impose the normalisation of the graviton kinetic terms,
\begin{equation}
\label{Gnorm}
\frac{M_{*}^3}{k}\int_0^Ldy~e^{-2ky}f_n^gf_m^g\left(k+r_0\delta(y)+r_L\delta(y-L)\right)=4\delta_{mn}.
\end{equation}
The normalised zero mode solution is then simply $f_0=2/M_{Pl}$ where,
\begin{equation}
\label{4DPl}
M_{Pl}^2=\frac{M_{*}^3}{k}\left( 1-e^{-2kL}+r_0+e^{-2kL}r_L \right)
\end{equation}
is the effective 4D Planck mass and determines the scale of the 4D gravity.  
In this work we will study how the phenomenology of the KK gravitons changes as we vary the 5D Planck scale ($\sim M_*$) and the 5D volume ($kL$).
In our work we fix $M_{KK}$ to some value meaning that $k$ and $L$ are not independent.
After fixing $M_{KK}$ we are left to vary $k/M_*$ and $kL$, where $k$ is fixed by the chosen value for $M_{KK}$.

An important point to make is that NDA perturbativity bounds imply that for $k\lesssim2M_*$ higher order curvature terms in the 5D action can comfortably be neglected \cite{Agashe:2007zd}. 
We will see that the couplings of the KK graviton are suppressed by $k/M_*$, and for $k\ll M_*$ the KK graviton bounds are very weak.
However since we only have one scale in our 5D model this would imply tuning.
To be completely general we will allow $k/M_*$ to vary in the range $[0.5,3]$.

Lowering $M_*$ inevitably means that the 4D Planck scale is also lowered.
To remedy this we make the kinetic term on the UV brane ($r_0$) very large such that it compensates for this reduction in $M_*$ and reproduces the correct 4D gravity scale\footnote{We would like to thank Andrea Wulzer for useful comments and discussions on this topic.}\cite{George:2011sw}.
As discussed previously in this section, such a large BKT imposes an effective UV Dirichlet boundary condition for the KK graviton modes.
A small UV scale $M_*$ also implies a small $k$, and if $M_{KK}$ is fixed this in turn implies a smaller value for the 5D volume $kL$, hence these models being referred to as {\it Little Randall-Sundrum} models \cite{Davoudiasl:2008hx,McDonald:2008ss,Davoudiasl:2008nr,Bauer:2008xb,Davoudiasl:2009jk,Davoudiasl:2010fb,Croon:2015wba}.

In these models, contributions to some flavour and electroweak precision observables are reduced due to a volume suppression $(\sim kL)$, although additional flavour symmetries may have to be invoked to prevent proton decay.  More recently it has been shown that holographic realisations of Composite Higgs models with this lower cutoff may require less fine-tuning in the Higgs potential in order to generate a light Higgs, yet compatible with the absence of light top-partners \cite{Croon:2015wba}.

It is interesting to note that lowering this UV scale in the 5D model corresponds to a reduction of the cutoff in the dual 4D gauge theory, and also an increase in the number of colours in the strongly coupled gauge theory.
The relationship is precisely
\beq
kL=\frac{4M_{KK}^2}{g^2f_{\pi}^2}=\frac{16\pi^2}{g^2N}
\eeq
where $f_{\pi}$ is the decay constant of the composite pseudo-Goldstone Higgs boson, $g$ is the weak coupling, and $N$ the number of colours. Despite recent LHC results, values of $f_{\pi}\gtrsim500$ GeV are still in agreement with data \cite{Sanz:2017tco}.

\subsection{Couplings to the Standard Model}

We assume that all SM fields are in the bulk of the extra dimension.
Bulk fields can be decomposed into towers of modes which may, depending on boundary conditions, have a massless zero mode.
We are only interested in these massless zero modes, since these are identified with the SM degrees of freedom, thus we drop the massive terms in the mode expansion from here on.
It is possible that other zero modes besides the SM states exist, however we do not consider these scenarios here.

In the holographic composite Higgs models the Higgs arises as the fifth component of a bulk gauge symmetry, which fixes its 5D wave function. 
The wave functions of the gauge fields are also fixed by gauge invariance.
The wave functions of the transverse components are flat, whereas the wave functions of the longitudinal components have the same localisation as the Higgs boson.
The fermion wave functions are different as we are free to localise these anywhere in the bulk.
However if we wish to explain the fermion mass hierarchies naturally in this scenario we require the top quark to be IR localised and the other quarks to be localised away from the IR brane.
In the case where the light SM fermions are localised towards the UV brane it is sufficient to only consider the third generation of quarks in the phenomenological analysis.

We can parameterise the wave functions of the SM fields in the extra dimension with
\begin{equation}
\label{fa}
f_a=\sqrt{\frac{2ak}{1-e^{-2akL}}}e^{-aky}
\end{equation}
where the kinetic terms of these fields are normalised to 1, i.e. $\int_0^Ldy~f_a^2=1$, and $a=(a_h$, $a_A$,$a_{q}$, $a_{tr}$, $a_{br})$ are the 5D mass/localisation parameters of the Higgs, transverse gauge fields, left-handed top doublet, right-handed top, and right handed bottom.
The Higgs and massless gauge field parameters are fixed to $a_h=-1$ and $a_A=0$.
Values of $a$ less than zero imply that the field is localised towards the IR brane, thus the Higgs in this case is IR localised.
We also expect $a_{tr}<0$ whereas we expect $a_q\sim 0$ and $a_{br}\gg0$.

The graviton couplings to these particles are then given by overlap integrals of the SM wave functions with the graviton wave function,
\begin{equation}
\label{ca}
c_a=\frac{v}{2}\int_0^Lf_a^2f_1^g dy.
\end{equation}
In the limit where $f_a\sim 1$ is flat along the extra dimension we find that
\begin{equation}
\label{cflat}
c_{flat}\simeq\left(\frac{k}{M_{*}}\right)^{3/2}\frac{m_Xv}{M_{KK}^2}\frac{1}{8kL}.
\end{equation}
Thus the couplings of the KK graviton to transverse gauge fields are strongly affected by varying $kL$, or in our case, varying the UV scale.
This feature is unique to fields with flat wave functions, it is not seen in the couplings to the Higgs or to fermions (unless the fermions have a flat wave function).
The effects of varying $kL$ contrast with varying $k/M_*$.
From eq.~\ref{Gnorm} we see that the kinetic terms have a $M_*/k$ pre-factor, thus upon normalisation the couplings of the KK gravitons to all matter will scale with $k/M_*$.

Although we are not interested in the effects of the KK modes in this work, their effects on electroweak precision observables are worth a short discussion.
Without a bulk custodial symmetry \cite{Agashe:2003zs} to protect the EWPOs from large corrections, the scale $M_{KK}$ is constrained to lay $\gtrsim 10$ TeV. 
For discussions on EWPOs in warped extra dimensions see \cite{Agashe:2003zs,Cabrer:2011fb,Dillon:2014zea}.
However such a large mass gap between the electroweak scale and the new physics scale seems unnatural and fine-tuned.
The bulk custodial symmetry is an elegant solution which enlarges the spectrum of KK modes and allows for the scale $M_{KK}$ to lay $\gtrsim 1$ TeV.
Most realistic minimal composite Higgs models also naturally generate this symmetry, for example the $SO(5)/SO(4)$ models \cite{Agashe:2004rs}.
In our work we will naturally assume that this symmetry is present and allow $M_{KK}$ to be at $1$ TeV.

\section{Production and Decay of the Little Graviton}

The graviton interactions with SM particles are given by dimension-five operators which we will normalize to the electroweak scale $v$ for convenience. The specific expressions of these operators can be found elsewhere in the literature, e.g.~\cite{Lee:2013bua}. One can then compute the partial decay widths of the graviton, which we will denote by $X_{\mu\nu}$, to the SM fields~\cite{Lee:2013bua,Falkowski:2016glr},
\begin{align}
    \Gamma(X\rightarrow gg)=& \frac{c_g^2m_X^3}{10\pi v^2}  \ , \,   \Gamma(X\rightarrow \gamma\gamma)=\frac{c_{\gamma\gamma}^2m_X^3}{80\pi v^2}  \nonumber \\
    \Gamma(X\rightarrow hh)=&\frac{c_h^2m_X^3}{960\pi v^2}(1-4r_h)^{5/2}   \nonumber \\
    \Gamma(X\rightarrow f\bar{f})=&\frac{N_c(c_{fl}^2+c_{fr}^2)m_X^3}{320\pi v^2} (1-4r_f)^{3/2}(1+8r_f/3)  \nonumber \\
    \Gamma(X\rightarrow ZZ)=&\frac{m_X^3}{80\pi v^2}\sqrt{1-4r_Z}\left(c_{ZZ}^2+\frac{c_h^2}{12}\right. \nonumber \\
    &\left.+\frac{r_Z}{3}\left(3c_h^2-20c_hc_{ZZ}-9c_{ZZ}^2\right) \right.\nonumber \\
    &\left.+2\frac{r_Z^2}{3}\left(7c_h^2+10c_hc_{ZZ}+9c_{ZZ}^2\right)\right)   \nonumber \\
    \Gamma(X\rightarrow WW)=&  \frac{m_X^3}{40\pi v^2}\sqrt{1-4r_W}\left(c_{W}^2+\frac{c_h^2}{12}\right.\nonumber\\
    &\left.+\frac{r_W}{3}\left(3c_h^2-20c_hc_{W}-9c_{W}^2\right) \right.\nonumber \\
    &\left.+2\frac{r_W^2}{3}\left(7c_h^2+10c_hc_{W}+9c_{W}^2\right)\right)   \nonumber \\
    \Gamma(X\rightarrow Z\gamma)=& \frac{c_{Z\gamma}^2m_X^3}{40\pi v^2}(1-r_Z)^3\left(1+\frac{r_Z}{2}+\frac{r_Z^2}{6}\right)  
\end{align}
where $c_{\gamma\gamma}=s_{\theta}^2c_W+c_{\theta}^2c_B$, $c_{ZZ}=c_{\theta}^2c_W+s_{\theta}^2c_B$, $c_{Z\gamma}=s_{\theta}c_{\theta}(c_W-c_B)$, $r_i=(m_i/m_X)^2$, and $m_X$ is the lightest KK graviton mass.  The precise values of the Wilson coefficients, $c_i$, can be calculated from eq.~\ref{ca}.

Since the transverse components of the gauge fields are flat, one automatically obtains that, $c_g=c_{\gamma\gamma}=c_W=c_B$.  As a result, the decays to $Z\gamma$ are automatically absent.
This can only be altered if we include BKTs for the gauge fields~\cite{Falkowski:2016glr}, however one must be careful as these can alter the tree-level corrections to the electroweak precision observables.
For our purposes it makes sense to assume these BKTs for the electroweak bosons are absent.

\begin{figure}[h!]
  \begin{center}
\includegraphics[scale=0.4]{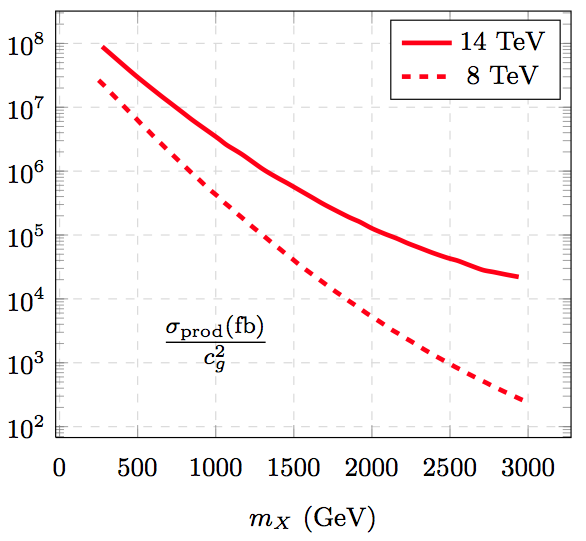}
    \caption{Production cross-section of the KK graviton via gluon fusion at centre of mass energies of $8$ and $14$ TeV.  The dependence of this quantity on the graviton coupling to gluons has been factored out. \label{Gxsec}}
  \end{center}
\end{figure}

The production cross-section for the KK graviton has been calculated at leading order in $pp\rightarrow G$ collisions at $s=8,14$ TeV assuming that gluon fusion is the dominant process \cite{Gouzevitch:2013qca}.
In figure~\ref{Gxsec} we plot the production cross-section for the KK graviton at the relevant centre-of-mass energies.
The dependence of this cross-section on the coupling to gluons is trivial and has thus been factored out.

\section{The experimental bounds}
In the plots below we have collected the $95\%$ CL bounds from heavy resonance searches at ATLAS an CMS.  
The decay products we are interested in are $gg\rightarrow XX\rightarrow\gamma\gamma, WW, ZZ, t\bar{t}, \text{ and }  hh$.
In each case we have looked at each available final state and taken the most stringent constraint at each mass point.

\begin{figure}[h!]
  \begin{center}
\includegraphics[scale=0.34]{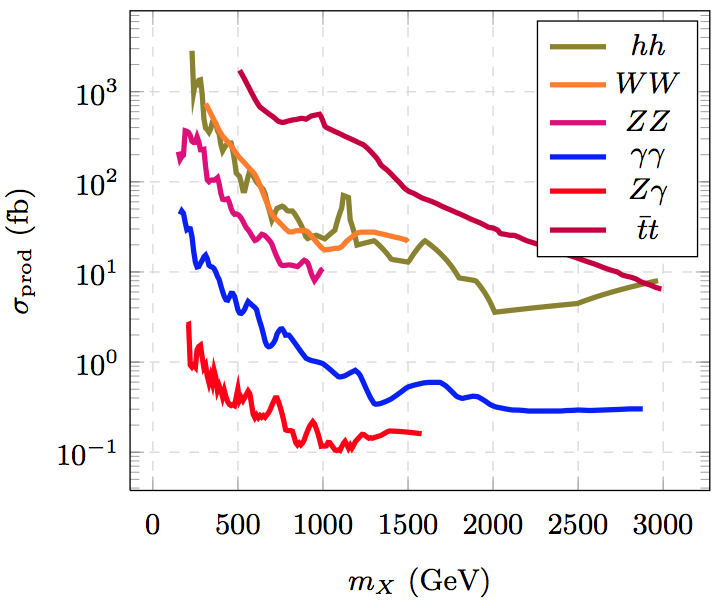}
\caption{Relevant bounds on the graviton mass from the ATLAS and CMS analyses of run 1 data \cite{Aad:2015mna,Aad:2015agg,Aad:2015kna,ATLAS:2014rxa,Aad:2014yja,Aad:2015fna,CMS:2014onr,Khachatryan:2016sey,Khachatryan:2015yea,Khachatryan:2015tha,CMS:2016zxv,CMS:2015zug,Khachatryan:2016cfa}.  \label{bounds1}}
  \end{center}
\end{figure}

\begin{figure}[h!]
  \begin{center}
\includegraphics[scale=0.34]{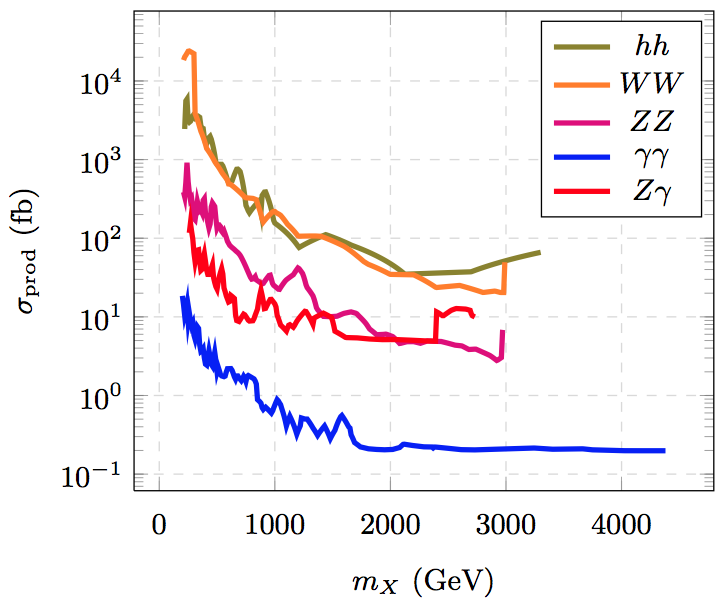}    
\caption{Bounds on the graviton mass from the ATLAS and CMS analyses of run 2 data \cite{ATLAS:2016eeo,ATLAS:2016kjy,ATLAS:2016cwq,ATLAS:2016oum,ATLAS:2016bza,ATLAS:2016npe,ATLAS:2016qmt,ATLAS:2016btu,Khachatryan:2016yec,CMS:2016jpd,CMS:2016ilx,CMS:2016rec,CMS:2016knm,CMS:2016tlj,CMS:2016pwo,CMS:2016vpz}. 
   \label{bounds2}}
  \end{center}
\end{figure}
Without graviton BKTs and with $M_{KK}\sim 1$ TeV (as required by electroweak precision bounds) the lightest KK mode is $\gtrsim 3.8$ TeV, thus the current bounds are not able to sufficiently probe this parameter space.
In this work we study the parameter space of light KK gravitons with masses in the range $[0.5,3]$ TeV, where bounds from most channels are relevant.

\section{A KK Graviton and a composite Higgs}
In this section we will present our results on studying the phenomenology of a KK graviton in a holographic composite Higgs model.
Our free parameters are $M_{KK}$, $k/M_*$, $kL$, the third generation wave functions, and $m_X$ (the graviton mass).
In performing our analysis we will keep $M_{KK}$ constant at $1$ TeV, and in each parameter scan we will keep the top quark localisations and $m_X$ constant while varying $kL=\log(\Omega)$ and $k/M_*$.
For the third generation wave functions we choose $a_{tr}=-0.3$ and $a_{q}=0$, and for the KK graviton masses we take $m_X=500$, $750$, $1000$, $1250$, $1500$, $2000$, $2500$, and $3000$ GeV.
In table \ref{masspxsec} we collect the production cross-sections of the KK graviton for each mass point at run 1 and run 2 of the LHC, and in table \ref{bktmx} we collect the values of the IR BKT required to generate the appropriate mass.
Note that the approximation of a large BKT in eq. \ref{mxrl} is only accurate for KK graviton masses below $1500$ GeV. 

\begin{figure*}[ht!]
  \begin{center}
\includegraphics[scale=0.67]{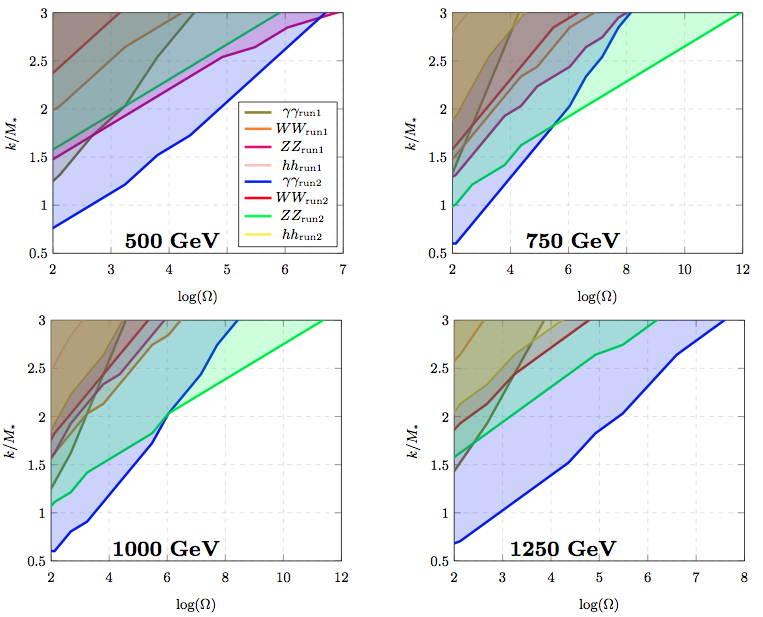} 
\caption{Exclusion bounds for a light KK graviton with masses of $500,~750,~1000$, and $1250$ GeV.\label{light}}
 \end{center}
 \end{figure*}

\begin{figure*}[ht!]
\begin{center}
\includegraphics[scale=0.67]{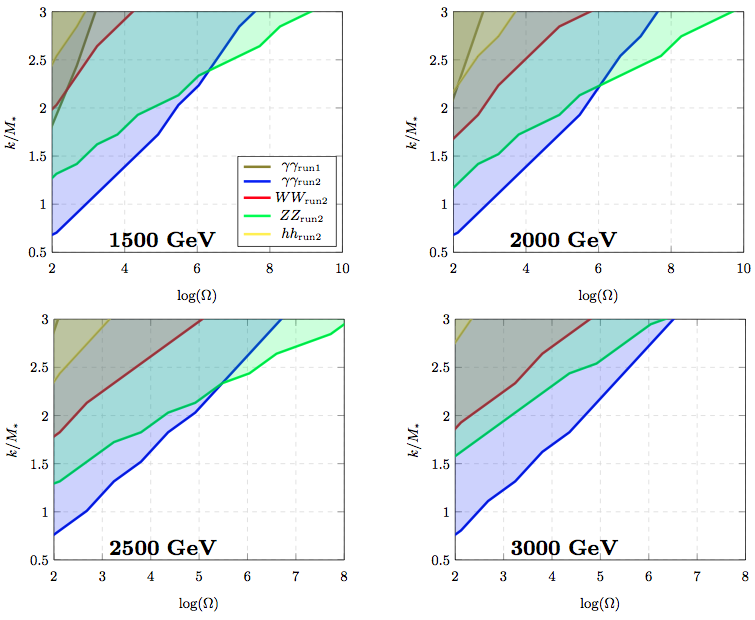} 
\caption{Exclusion bounds for a heavier KK graviton with masses of $1500,~2000,~2500$, and $3000$ GeV.\label{heavy}}
  \end{center}
\end{figure*}

We split the masses into two categories; the light KK gravitons ($m_X=500\ldots1250$ GeV) and the heavier KK gravitons ($m_X=1500\ldots3000$ GeV). 
We will first discuss the lighter states, for which the exclusion plots are shown in figure \ref{light}.
In most of the parameter space the run 2 di-photon bounds are the most constraining, except for large $\log(\Omega)$ values at masses of $750$ GeV and $1000$ GeV, where the $ZZ$ bounds dominate.
We see that lighter masses are not necessarily more constrained than heavier masses, as one would usually expect when the KK graviton mass is altered by varying $M_{KK}$.
This is because the couplings of the KK graviton are significantly enhanced for smaller values of $M_{KK}$, however in our work we keep $M_{KK}$ constant and reduce the KK graviton mass using an IR BKT.
In this sense the phenomenology discussed here is quite different from the usual KK graviton studies.
This is more apparent in the heavier mass ranges whose exclusion bounds are plotted in figure \ref{heavy}.
Here we see that the heavier masses are generally no less constrained than the lighter masses.
We also see similar behaviour in the di-photon and $ZZ$ channels providing the strongest bounds.
The di-photon bound always dominates at lower $\log(\Omega)$ values.
This is due to the enhancement of the photon coupling to the KK graviton at low $\log(\Omega)$ as shown in eq. \ref{cflat}.
If the 5D Higgs wavefunction was flat then the $Z$ and $W$ couplings would also be enhanced at low volume, but the fact that we have a pseudo-Goldstone Higgs fixes its IR localisation.

The free parameters that we have not varied here are $M_{KK}$ and the top localisations.
We have kept $M_{KK}$ at $1$ TeV such that electroweak precision tests are satisfied.  Increasing this parameter would simply lead to a global increase in the KK graviton masses and a global reduction in the KK graviton couplings.
The top localisations do not play a major role in constraining the parameter space.
We can see from figures \ref{light} and \ref{heavy} that the bounds from the top decays are negligible in comparison to the principal decay modes.
The top couplings to the KK graviton would be enhanced by moving the left or right handed wavefunctions towards the IR.
However to reproduce the correct bottom quark mass the left handed localisation cannot be too IR localised, and the right handed localisation that we have chosen in our computations is already sufficiently IR localised.
Increasing this does not significantly affect the results.

The main result from our study is that light KK gravitons arising in holographic composite Higgs models are not ruled out by direct detection bounds at the LHC.
It is also clear that the channels most sensitive to the presence of these states are the di-photon and $ZZ$ final states.
Thus any signal from this model would be expected to first show up in one of these channels.
A signal in the di-photon channel would indicate a KK graviton from a 5D model with a small 5D volume, or a 4D strongly coupled gauge theory with a large number of colours.
Whereas a signal in the $ZZ$ channel would indicate a larger 5D volume, or a smaller number of colours in the dual 4D gauge theory.

\begin{table}
\centering
\begin{tabular}{lll}
   $m_X$ (GeV) &\vline~Run1 (fb/$c_g^2$)  & Run2 (fb/$c_g^2$) \\ \hline
500&\vline~$6.3\times10^6$ & $1.6\times10^7$ \\
750&\vline~$1.4\times10^6$ & $1.1\times10^7$ \\
1000&\vline~$4.0\times10^5$ & $3.3\times10^6$ \\
1250&\vline~$1.2\times10^5$ & $1.3\times10^6$ \\
1500&\vline~$3.9\times10^4$ & $5.6\times10^5$  \\
2000&\vline~$5.3\times10^3$ & $1.3\times10^5$ \\
2500&\vline~$9.9\times 10^2$  & $4.3\times 10^4$  \\
3000&\vline~$2.4\times 10^2$  & $2.1\times 10^4$  \\
\end{tabular}
\caption{Values of the production cross-sections of the KK graviton from figure~\ref{Gxsec} for specific values of the graviton mass. \label{masspxsec}}
\end{table}

\begin{table}
\centering
\begin{tabular}{lll}
   $m_X$ (GeV) &\vline~$r_L$  \\ \hline
500&\vline~ 15.8 \\
750&\vline~ 6.95 \\
1000&\vline~ 3.8 \\
1250&\vline~ 2.4 \\
1500&\vline~ 1.6  \\
2000&\vline~ 0.8 \\
2500&\vline~ 0.45     \\
3000&\vline~ 0.23     \\
\end{tabular}
\caption{List of the BKT values required to generate the relevant KK graviton mass. \label{bktmx}}
\end{table}

\begin{figure}[h!]
  \begin{center}
\includegraphics[scale=0.4]{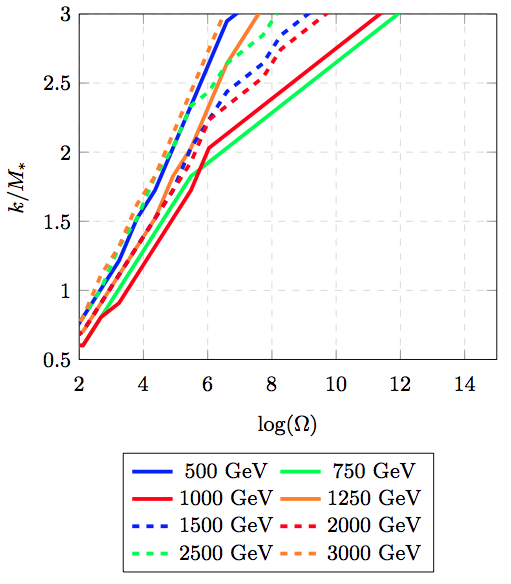}    
\caption{Comparison of exclusion bounds for different KK graviton masses.  The parameter space above each curve is excluded in each case.\label{scan1summary}}
  \end{center}
\end{figure}

\section{Conclusions}

In this paper we used 5D holographic techniques to study experimental bounds on a spin-2 composite state (KK graviton) arising specifically in a composite Higgs model.
We considered masses for the spin-2 state (KK graviton) in the range $500-3000$ GeV, however unlike in most studies of these scenarios we did so by keeping the KK scale constant and varying the size of a BKT on the IR brane to fix the KK graviton mass.
Keeping this scale fixed at $1$ TeV we work in a regime were corrections to EWPOs from heavy spin-1 states are under control.
Having no IR BKT for the graviton would imply that the lightest spin-2 state has a mass of $3.8$ TeV, and thus lays out of reach of all heavy resonance searches except those in the di-photon channel.
With moderate values of the IR BKT this mass can be significantly reduced, bringing it well within the reach of the LHC, therefore its phenomenology is certainly worth studying.

With the 5D localisation of the Higgs wave function fixed, and the insensitivity of experimental bounds to the top localisations, the interesting parameter space in this model consists of the 5D volume $\log(\Omega)$ and the ratio $k/M_*$.
We presented exclusion plots in these variables for each mass point we studied and found that, even for the lighter KK gravitons, much of the parameter space is unconstrained.
The reason the parameter space for these light states remains largely unconstrained is because the smallness of their masses is generated by the IR BKT rather than the KK scale.
If we were to lower the KK scale to suppress the KK graviton masses then the couplings to the SM particles would increase, and the allowed parameter space would be reduced.
However because values of $M_{KK}$ less than $1$ TeV are ruled out by EWPO data we do not explore this region.
The most constraining searches are undoubtedly the $\gamma\gamma$ and $ZZ$ searches from run 2 of the LHC.
In particular, $\gamma\gamma$ bounds dominate at small 5D volumes (large number of colours in 4D confining gauge theory) and $ZZ$ bounds dominate at large 5D volumes (small number of colours in 4D confining gauge theory).

To conclude, using holographic methods we have shown that despite the stringent constraints imposed by the most recent LHC searches, spin-2 states with masses in range $500-3000$ GeV (a strong prediction in any composite Higgs scenario) are not excluded by experimental data.
The most likely channels for detecting one of these states are the $\gamma\gamma$ and $ZZ$ channels, where a signal first showing up in $\gamma\gamma$ would imply a large number of colours in the confining gauge theory and a signal in the $ZZ$ channel would indicate a smaller number of colours.

\section*{Acknowledgements}
BMD was supported by the Science and Technology Facilities Council (UK) in the early stages of this project and in later stages by EPSRC Grant EP/P005217/1. BMD would like to thank A.~Ahmed, B.~Grzadkowski, J.~F.~Gunion and Y.~Jiang, and A.~Wulzer for useful discussions.  VS would like to thank H.~M.~Lee and M.~Park for their long-standing collaboration on aspects of spin-2 resonances in Beyond the Standard Model.  Both authors would like to thank Kaustubh Agashe for illuminating discussions on the nature of the radion in the presence of brane localised gravity terms.

\section*{References}

\bibliography{KKgravitonlast}

\end{document}